\documentclass[aps,prl,twocolumn,superscriptaddress]{revtex4}
\usepackage{txfonts}
\usepackage{graphicx}
\usepackage{color}

\begin{document}

\title{Observation of Majorana Fermions in a Nb-InSb Nanowire-Nb Hybrid Quantum Device}

\author{M. T. Deng}
\affiliation{Division of Solid State Physics, Lund University, Box 118, S-221 00 Lund, Sweden}
\author{C. L. Yu}
\affiliation{Division of Solid State Physics, Lund University, Box 118, S-221 00 Lund, Sweden}
\author{G. Y. Huang}
\affiliation{Division of Solid State Physics, Lund University, Box 118, S-221 00 Lund, Sweden}
\author{M. Larsson}
\affiliation{Division of Solid State Physics, Lund University, Box 118, S-221 00 Lund, Sweden}
\author{P. Caroff}
\affiliation{I.E.M.N., UMR CNRS 8520, Avenue Poincar\'{e},  BP 60069, F-59652 Villeneuve d'Ascq, France}
\author{H. Q. Xu}
\email[Corresponding author: ]{hongqi.xu@ftf.lth.se}
\affiliation{Division of Solid State Physics, Lund University, Box 118, S-221 00 Lund, Sweden}
\affiliation{Department of Electronics and Key Laboratory for the Physics and Chemistry of Nanodevices, Peking University, Beijing 100871, China}

\date{March 27, 2012}

\begin{abstract}
\textbf{The search for Majorana fermions is one of the most prominent research tasks in modern physics \cite{MFreturns,Franz10,MFnearSuccess,Beenakker11,MFinSolid}. Majorana fermions are an elusive class of fermions that act as their own antiparticles~\cite{Majorana}. However, the fundamental nature of Majorana fermions can be easily viewed using the second quantization notation in which the creation of a fermion such as an electron can be described by the excitation of two Majorana fermions \cite{Beenakker11,MFinSolid}. Although an extensive effort has been made worldwide in particle physics, Majorana fermions have so far not been convincingly discovered in free space. In recent years, numerous proposals~\cite{KitaevHybridized,5-2FQHEasMF,NonAbelian,ReadGreenPwave,FuKaneSwaveTi,JaySauMFinSemi,AliceaMFin2DSemi,SauPRB10,DSarmaMFin1DSemi,OregHelicalMF,SDS-PRB,MF1DNetwork,FlensbergMFQD2} for excitations of Majorana fermions in solid state systems have been made, ranging from exploring $\nuup =5/2$ fractional quantum Hall systems~\cite{5-2FQHEasMF,NonAbelian}, to exploring chiral p-wave superconductors~\cite{ReadGreenPwave}, and to exploring hybrid systems of a topological insulator~\cite{FuKaneSwaveTi} or a strong spin-orbit coupled semiconductor thin film~\cite{JaySauMFinSemi,AliceaMFin2DSemi,SauPRB10} or nanowire~\cite{DSarmaMFin1DSemi,OregHelicalMF,SDS-PRB} in the proximity of an external s-wave superconductor. These proposals  have stimulated a new wave of search for Majorana fermions~\cite{MFreturns,Franz10,MFnearSuccess}. Here, we report on the observation of excitation of Majorana fermions in a Nb-InSb nanowire quantum dot-Nb hybrid system. The InSb nanowire quantum dot is formed between the two Nb contacts by weak Schottky barriers and is thus in the regime of strong couplings to the contacts. Due to the proximity effect, the InSb nanowire segments covered by superconductor Nb contacts turn to superconductors with a superconducting energy gap $\Delta^*$. Under an applied magnetic field larger than a critical value for which the Zeeman energy in the InSb nanowire is $E_z\sim \Delta^*$, the entire InSb nanowire is found to be in a nontrivial topological superconductor phase, supporting a pair of Majorana fermions, and Cooper pairs can transport between the superconductor Nb contacts via the Majorana fermion states. This transport process will be suppressed when the applied magnetic field becomes larger than a second critical value at which the transition to a trivial topological superconductor phase occurs in the system. This physical scenario has been observed in our experiment. We have found that the measured zero-bias conductance for our hybrid device shows a conductance plateau in a range of the applied magnetic field in quasi-particle Coulomb blockade regions. This work provides a simple, solid way of detecting Majorana fermions in solid state systems and should greatly stimulate Majorana fermion research and applications.}
\end{abstract}

%\pacs{XXXXXX}

\maketitle

Epitaxially grown InSb nanowires are a class of the most promising material systems for the formation of a hybrid device with an s-wave superconductor in which the excitation of Majorana fermions can be achieved under an application of an external magnetic field of a moderate strength. InSb nanowires~\cite{Giant-gfactor,Henrik-PRL} possess a large electron $g$ factor ($|g^*|\sim 30-70$), a strong spin-orbit interaction strength (with a spin-orbit interaction energy in the order of $\Delta_{SOI}\sim 0.3$ meV), and a small electron effective mass ($m^*\sim 0.015 m_e$). These properties allow us to generate a helical liquid in the InSb nanowire by applying a relatively small magnetic field and to generate a nontrivial topological superconductor~\cite{DSarmaMFin1DSemi,OregHelicalMF,SDS-PRB}, which supports a pair of Majorana fermions, by coupling the InSb nanowire to an s-wave superconductor in an experimentally feasible condition. The s-wave superconductor will introduce superconductivity into the InSb nanowire by proximity effect~\cite{Henrik-Super} and the external magnetic field will then drive the strongly spin-orbit coupled nanowire system into a nontrivial topological superconductor phase through the Zeeman splitting. The giant Land\'{e} g-factor of the InSb nanowire~\cite{Giant-gfactor,Henrik-PRL} guarantees a significantly large Zeeman splitting at a magnetic field well below the critical magnetic field of the superconductor.

\begin{figure}
\includegraphics[width=8.5cm]{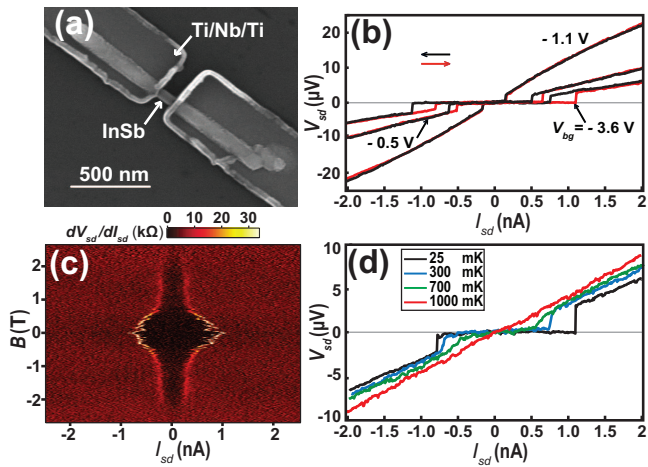}
\caption{\label{fig:1} \textbf{Device layout and Josephson current.} (a) SEM image of the Nb-InSb nanowire-Nb junction device studied in this work. The device is made from the zincblende InSb segment of an InAs/InSb heterostructure nanowire with two Ti/Nb/Ti (3 nm/80 nm/5 nm) contacts on a SiO$_2$ capped, highly doped Si substrate, using electron beam lithography, sputtering, and lift-off process. The diameter of the InSb nanowire is about 65 nm, the separation between the two Nb based contacts is about 110 nm, and the lengths of the InSb nanowire sections covered by the two Nb based contacts are about 740 nm and 680 nm, respectively. (b) Source-drain voltage $V_{sd}$ measured for the device as a function of source-drain current $I_{sd}$ at three different back gate voltages $V_{bg}=-0.5$, $-1.1$ and $-3.6$ V and at temperature $T=25$ mK. The red and black curves are recorded in the upward and the downward current sweeping direction, respectively. Hysteresis is seen in the $V_{sd}-I_{sd}$ characteristics of the device measured at $V_{bg}=-0.5$ and $-3.6$ V.  A supercurrent with the critical value $I_c$ depending on $V_{bg}$ and the current sweeping direction is seen to flow through the InSb nanowire junction  due to the superconductivity induced by the proximity effect. (c) Differential resistance $dV_{sd}/dI_{sd}$ on a color scale measured for the device at $V_{bg}=-3.6$ V and $T=25$ mK as a function of source-drain current $I_{sd}$ and magnetic field $B$ applied perpendicularly to the substrate and thus to the nanowire. The supercurrent is seen to persist as the applied magnetic field goes beyond 2 T. (d) Source-drain voltage $V_{sd}$ measured for the device as a function of the source-drain current $I_{sd}$ at $V_{bg}=-3.6$ V and $B=0$ T and at four different temperatures $T=25$, $300$, $700$, and $1000$ mK. The supercurrent is seen to be still present at $T=700$ mK, but to disappear at $T\sim 1$ K.}
\end{figure}

A nontrivial topological superconductor could be achieved by covering an entire InSb nanowire with an s-wave superconductor forming an ``$\Omega$'' shaped superconductor contact. However, two technical difficulties for creating and detecting Majorana states may arise. First, it becomes difficult for the applied magnetic field to penetrate through the superconductor contact due to the Meissner effect. To get a sufficiently large Zeeman splitting, a strong magnetic field has to be applied, leading to the suppression of the superconductivity in the superconductor. Second, it is also technically difficult to probe the Majorana fermion states in such a  superconductor-wrapped nanowire by transport measurements. One way of solving the difficulties is to lay down an InSb nanowire on the surface of an s-wave superconductor and detect Majorana fermions in a scanning tunneling microscopy setup. However, to perform such an experiment at an ultra-low temperature is very challenging. In addition, it has not been experimentally proven that it is possible to achieve a strong coupling for introducing a sufficiently large proximity effect into the nanowire in this way. To solve these problems, we employ an InSb nanowire quantum dot based Josephson junction with two Nb contacts as shown in Fig.~\ref{fig:1}a. Here, Nb is employed because it has a high superconducting critical magnetic field~\cite{Hc2OfNb} and therefore a strong magnetic field can be applied to achieve a sufficiently large penetration field on the InSb nanowire. A most important advantage of using such a Josephson junction configuration is that the entire InSb nanowire can be turned to a coherent, nontrivial topological superconductor by proximity effect and, when Majorana fermion states appear in the InSb nanowire, we can detect them simply with use of the two Nb contacts in a standard transport measurement setup.

Our Nb-InSb nanowire quantum dot-Nb hybrid device is fabricated from the high crystalline quality, zincblende InSb segment of an epitaxially grown InSb/InAs heterostructure nanowire (see Methods for the details of the device fabrication). Figure~\ref{fig:1}a shows a scanning electron microscope (SEM) image of the fabricated device. The fabricated device is first characterized in a relatively open conduction region in which the conductance of the junction in the normal state is about $2e^2/h$ or higher. Figure~\ref{fig:1}b shows the measured source-drain voltage of the device at a temperature of 25 mK as a function of applied source-drain current $I_{sd}$ at three different voltages $V_{bg}$ applied to the back gate. A zero resistance branch is clearly seen in each measured curve, which indicates the presence of a dissipationless Josephson supercurrent in the junction. The Josephson junction switches to a dissipative transport branch when the applied current is larger than a critical value $I_{c}$. The upward current sweeping trace (red curve) and downward sweeping trace (black curve) have different switching points, i.e., the device shows a hysteretic behavior. This hysteretic behavior, which has also been seen in Josephson junctions made from semiconductor nanowires and superconductor Al~\cite{Henrik-Super,Doh05,Xiang06}, could be the result of phase instability typically found in a capacitively and resistively shunted Josephson junction or simply due to a heating effect. The supercurrent $I_{c}$ is related to the resistance of the junction at the normal state~\cite{Henrik-Super,Doh05,Xiang06} and can thus be tuned in our device using the back gate voltage $V_{bg}$. For example, in Fig.~\ref{fig:1}b, we see that $I_{c}$ is 1.1 nA at $V_{bg}=-3.6$ V and it is only 0.2 nA at $V_{bg}=-1.1$ V.

Figure~\ref{fig:1}c shows the measured differential resistance of our hybrid device at a fixed back gate voltage of $V_{bg}=-3.6$ V as a function of applied source-drain current $I_{sd}$ and magnetic field $B$ applied perpendicularly to the substrate and thus to the nanowire. It is generally seen that the supercurrent $I_c$ decreases as the magnetic field $B$ increases and disappears after the magnetic field becomes higher than a critical value $B_{c}$. In Al based Josephson junctions made from semiconductors, $B_{c}$ is generally found to be in a range of a few mT to a few 100 mT~\cite{Henrik-Super,Doh05,Xiang06}. Here we find that, with the Nb contacts, a value of $B_c > 2$ T can be achieved. Figure~\ref{fig:1}d displays the temperature evolution of the Josephson supercurrent $I_c$ of our device at $V_{bg}=-3.6$ V and $B=0$ T. As the temperature increases,  $I_c$ decreases gradually and eventually disappears when the temperature becomes higher than a critical value of $T_c\sim 1$ K.

\begin{figure}
\includegraphics[width=8.5cm]{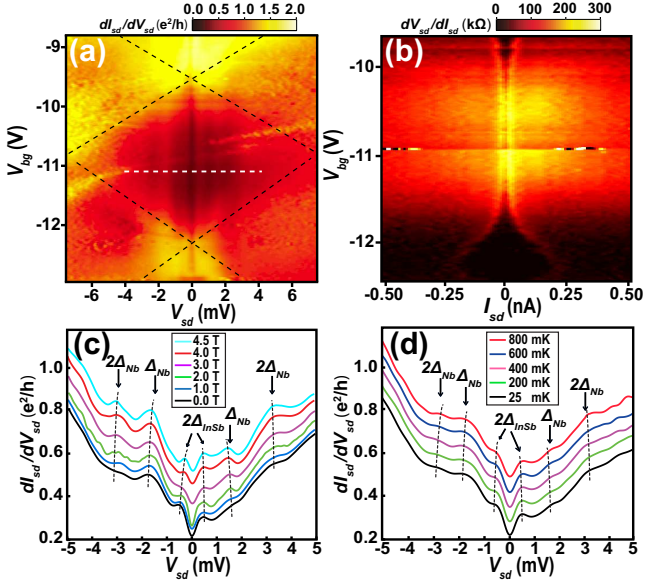}
\caption{\label{fig:2} \textbf{Superconductivity characteristics in a quasi-particle Coulomb blockade region.} (a) Differential conductance on a color scale as a function of source-drain voltage $V_{sd}$ and back gate voltage $V_{bg}$ (charge stability diagram) measured for the device at temperature $T=25$ mK and magnetic field $B=0$ T. The measurements show a Coulomb blockade diamond structure with a quasi-particle addition energy $\Delta E_{add} \sim 7$ meV to the InSb nanowire quantum dot in the device. The weak vertical lines of high conductance seen in the Coulomb blockade region arise from multiple Andreev reflections with the superconductor energy gaps, $\Delta_{Nb}$ and $\Delta_{InSb}$, related to both the two Nb based contacts and the two proximity effect induced superconducting InSb nanowire segments covered by the Nb based contacts. (b) Differential resistance on a color scale as a function of $I_{sd}$ and $V_{bg}$ measured for the device in the same Coulomb blockade region as in (a) at $T=25$ mK and $B=0$ T. A trace of the supercurrent characterized by a low resistance gap at $I_{sd}\sim 0$ nA is clearly seen in the Coulomb blockade region. (c) Differential conductance as a function of $V_{sd}$ measured at $V_{bg}=-11.1$ V, i.e., along the dashed line cut in (a), and $T=25$ mK, and at different magnetic fields applied perpendicularly to the substrate and thus to the InSb nanowire. The measured curves are successively offset by $0.02\, e^2/h$ for clarity. Here, multiple Andreev reflection induced conductance peaks due to the presence of the superconductor energy gaps $\Delta_{Nb}$ and $\Delta_{InSb}$ can be identified and are labeled for clarity. The dashed lines that mark the peak positions are used as guides to the eyes. (d) Differential conductance as a function of $V_{sd}$ measured at $V_{bg}=-11.1$ V, corresponding to the dashed line cut in (a), and $B=0$ T, and at different temperatures. Here, the measured curves are successively offset by $0.06\, e^2/h$ for clarity. The multiple Andreev reflection induced conductance peaks are again seen in the measurements and are labeled in the figure. The dashed lines are used once more as guides to the eyes.}
\end{figure}

Figure~\ref{fig:2} shows the measurements of our Josephson junction device in a low conduction region in which the quasi-particle transport shows the Coulomb blockade characteristics at low temperatures. Figure~\ref{fig:2}a displays the differential conductance of the device measured at $T=25$ mK and $B=0$ T as a function of source-drain bias voltage $V_{sd}$ and back gate voltage $V_{bg}$ (quasi-particle charge stability diagram). Here, a clear Coulomb blockade diamond structure is seen, indicating the formation of a quasi-particle quantum dot in the InSb nanowire junction region. This quasi-particle quantum dot is defined by the Schottky barriers present at the interfaces between the InSb nanowire and the two superconductor contacts. From the Coulomb diamond structure, we can determine that the quasi-particle addition energy of the quantum dot is $\sim 7$ meV, which is much larger than the superconductor energy gap $\Delta_{Nb}$ of Nb. Inside the Coulomb blockade region, vertical bright lines can be identified. These bright lines are due to multiple Andreev reflections and will be investigated further below. Figure~\ref{fig:2}b shows the differential resistance of the device measured at $T=25$ mK and $B=0$ T as a function of applied source-drain current $I_{sd}$ and back gate voltage $V_{bg}$ in the same quasi-particle Coulomb blockade region. Here, a distinct low resistance stripe at small $I_{sd}$ is clearly seen, indicating that the Cooper pair transport through the junction remains coherent. Such a coherent Cooper pair transport could arise from high-order co-tunneling processes~\cite{Dam06}.

As we mentioned above, the characteristics of multiple Andreev reflections are visible in the quasi-particle Coulomb blockade region. We will now investigate these characteristics in more details. Figure~\ref{fig:2}c shows the differential conductance as a function of $V_{sd}$ at a fixed back gate voltage of $V_{bg}=-11.1$ V, corresponding to the dashed line cut in Figure~\ref{fig:2}a, and $T=25$ mK, but for different perpendicularly applied magnetic fields. Here, we can, in each measured curve, clearly observe six conductance peaks, roughly located symmetrically around $V_{sd}=0$ V.  The four peaks at large $|V_{sd}|$ arise from the first and second order multiple Andreev reflections associated with the superconductor energy gap $\Delta_{Nb}$ of the Nb contacts as indicated in the figure. The two peaks at small $|V_{sd}|$ can be interpreted as arising from the first order multiple Andreev reflection associated with the proximity effect induced superconductor energy gap $\Delta_{InSb}$ of the InSb nanowires covered by the Nb contacts~\cite{Nb-InGaAs-PRB}. The $|V_{sd}|$ positions of these peaks also show weak dependences on the applied magnetic field in the range of $0\sim 4.5$ T, indicating that the values of the critical magnetic field $B_c$ in the thin Nb contacts are large and go beyond 4.5 T. The measurements also imply that the proximity effect induced superconductivity in the Nb-wrapped InSb nanowire segments has not been completely destroyed by the perpendicularly applied magnetic field of strength up to 4.5 T. However, we should note that due to the Meissner effect, the magnetic field penetrating through the Nb contacts and reaching the Nb-wrapped InSb nanowire segments is one order smaller than the applied magnetic field. Thus, the critical magnetic field of the proximity effect induced InSb nanowire superconductors should be in the order of a few 100 mT. Figure~\ref{fig:2}d shows the measured differential conductance as a function of $V_{sd}$ at $V_{bg}=-11.1$ V and $B=0$ T, but at different temperatures.  Here, we observe again that weak temperature dependences of the $|V_{sd}|$ positions of the multiple Andreev reflection conductance peaks, indicating that the values of $T_c$ are high and go beyond 1 K in the Nb and the proximity effect induced InSb nanowire superconductors. Furthermore, from the measured multiple Andreev reflection conductance peaks shown in Figs.~\ref{fig:2}c and \ref{fig:2}d, we can deduce ${\Delta _{Nb}} = 1.55$ meV and ${\Delta _{InSb}} = 0.25$ meV.

\begin{figure}
\includegraphics[width=8.5cm]{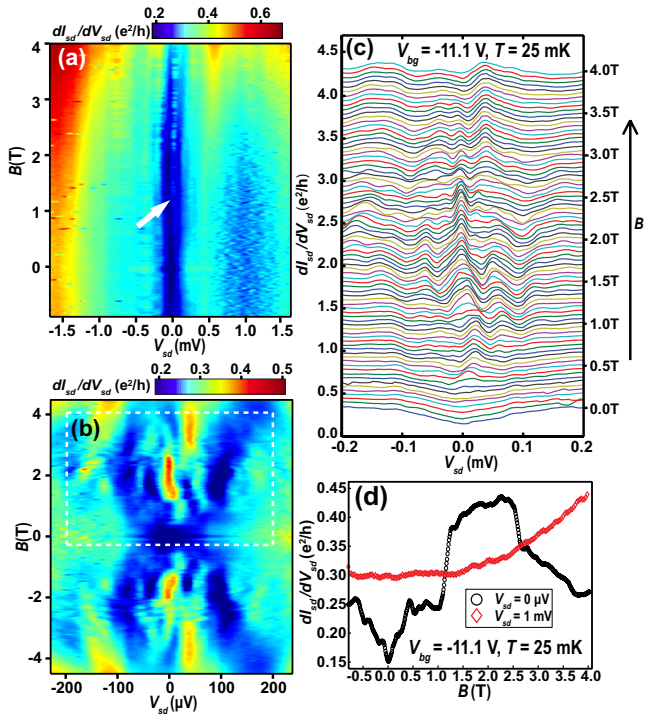}
\caption{\label{fig:3} \textbf{Characteristics of the transport through Majorana fermion states.} (a) Differential conductance on a color scale as a function of source-drain bias voltage $V_{sd}$ and magnetic field $B$ applied perpendicularly to the substrate and thus to the InSb nanowire measured at $V_{bg}=-11.1$ V (corresponding to the dashed line cut in Fig.~\ref{fig:2}a) and $T=25$ mK. A pronounced zero bias conductance peak (indicated by a white arrow) as a result of the Cooper pair transport through a pair of Majorana fermion states in the nontrivial topological superconductor InSb nanowire is seen to emerge when the magnetic field is above 1 T. (b) The same measurements as in (a) but with a high magnetic field resolution. Here, the zero-bias conductance peaks are more clearly seen and appear in both the positive and the negative magnetic field direction. (c) The same measurements as in (b) but the differential conductance traces measured in the region marked by a dashed square in (b) are shown. Here, the measured curves are successively offset by $0.05\, e^2/h$ for clarity. In this figure, accompanied to the zero-bias conductance peaks, two side differential conductance peaks are seen to appear at finite $V_{sd}$. (d) Differential conductance as a function of $B$ at $V_{sd}=0$ $\mu$V (black circles) corresponding to a line plot along $V_{sd}=0$ $\mu$V in (c).  Here, it is seen that the enhanced zero-bias conductance by the transport through the Majorana fermion states shows a plateau structure. The slope in the plateau could be due to a change in the background conductance with increasing $B$ in the region, as indicated by the differential conductance measurements at $V_{sd}=1$ mV (red diamonds).}
\end{figure}

To search for Majorana fermion states, we drive the Nb wrapped InSb nanowire segments from a trivial superconductor phase to a nontrivial topological superconductor phase. This is done by applying an external magnetic field perpendicular to the nanowire (and also to the substrate). The magnetic field introduces a Zeeman energy ${E_z} = \frac{1}{2}\left| {g^*} \right|{\mu _B}\tilde{B}$, where ${\mu _B} = {{e\hbar } \mathord{\left/ {\vphantom {{e\hbar } {2{m_e}}}} \right.\kern-\nulldelimiterspace} {2{m_e}}}$ is the Bohr magneton, $g^*$ is the effective g-factor, and $\tilde{B}$ is the magnetic field actually applied on the Nb wrapped InSb nanowire segments. The phase transition from the trivial superconductor phase to the nontrivial topological superconductor phase in the Nb wrapped InSb nanowire segments occurs at ${E_z} = \sqrt {\Delta _{InSb}^2 + {\mu ^2}}$, where $\mu $ is the chemical potential~\cite{JaySauMFinSemi}. It is difficult to accurately determine the strength of the externally applied magnetic field $B_T$ at which the phase transition in the InSb nanowire occurs. However, using the Nb bulk value of $\lambda_L\sim 390$ \AA \ for the magnetic field penetration depth~\cite{Nb-Detpth} and assuming the $g$-factor $|g^*|\sim 30-70$ and the superconductor energy gap ${\Delta_{InSb}}\sim 0.25$ meV for the InSb nanowire, we can estimate that the externally applied magnetic field that can induce the phase transition in the Nb wrapped InSb nanowire segments is in the range of $1-2$ T.

When there is no coupling between the two Nb wrapped InSb nanowire segments, each of the two nanowire segments in the nontrivial topological superconductor phase supports a pair of Majorana fermions located at the two ends of the segment~\cite{FlensbergMFQD2}. When the two nontrivial topological superconducting InSb nanowire segments are coherently connected via a normal state quantum object, e.g., quantum dot, the interaction between a pair of nearby zero energy Majorana fermions leads to creation of a pair of fermion states at finite energies, as a result of annihilation of the pair of Majorana fermions, and the entire nanowire will turn to a nontrivial topological superconductor supporting the pair of the remaining zero energy Majorana fermions (see Supplementary Information for the results of our modeling and simulation), which are now located at the two ends of the entire nanowire and are therefore strongly coupled to the Nb contacts. Cooper pairs can transport between the two Nb contacts via the Majorana fermion states, leading to an enhancement in the zero-bias conductance~\cite{MF1DNetwork, KitaevHybridized}. We will show below that such an enhancement in the zero-bias conductance has been observed in our Nb-InSb nanowire quantum dot-Nb device in the quasi-particle Coulomb blockade regime.

We first present our magnetic field dependent measurements of the differential conductance at the back gate voltage $V_{bg}=-11.1$ V for which the InSb nanowire quantum dot is in a Coulomb blockade low conductance region. Figure~\ref{fig:3}a shows the results of the measurements. Here, a low conductance gap can be clearly identified in the small source-drain bias voltage region. This low conductance gap structure can be attributed to the presence of the superconductor energy gap $\Delta _{InSb}$ in the Nb wrapped InSb nanowire segments. Most importantly, Fig.~\ref{fig:3}a shows, as expected, the emergence of a pronounced high zero-bias conductance structure, indicated by the white arrow in the figure, as the magnetic field exceeds  1.2 T. To see this high zero-bias conductance structure more clearly, we show in Fig.~\ref{fig:3}b high-resolution magnetic field dependent measurements of the differential conductance at $V_{bg}=-11.1$ V. Here, two high zero-bias conductance structures, located symmetrically with the respect to the zero magnetic field, can be clearly identified in the magnetic field regions with strengths around $2$ T. Figure~\ref{fig:3}c shows close-up traces for visualization of the details in the region of the measurements marked by a dashed square in Fig.~\ref{fig:3}b. Here, we see that the differential conductance shows rich  structures. In particular, two side differential conductance peaks emerge along with the zero-bias conductance peak and move apart in source-drain bias voltage as the applied magnetic field is increased. 
Figure~\ref{fig:3}d shows a trace of the differential conductance at the zero bias voltage, where the high zero-bias conductance structure is seen to appear as a plateau. Figure~\ref{fig:3}d also shows a corresponding trace of the differential conductance at a finite bias voltage of $V_{sd}=1$ mV. A comparison between the two traces may imply that the small slope seen in the zero-bias conductance plateau is due to a change in the background differential conductance with increasing $V_{bg}$. This high zero-bias conductance plateau structure is fully consistent with our physical picture that a pair of Majorana fermions can be excited in a certain range of the applied magnetic field for which the entire InSb nanowire is in a nontrivial topological superconductor phase. In addition, This high zero-bias conductance plateau structure implies that the transport through a pair of Majorana fermions will contribute to the conductance with an approximately constant value (which may strongly depend on the phase difference between the two superconductor Nb contacts~\cite{KitaevHybridized}).

\begin{figure}
\includegraphics[width=8.5cm]{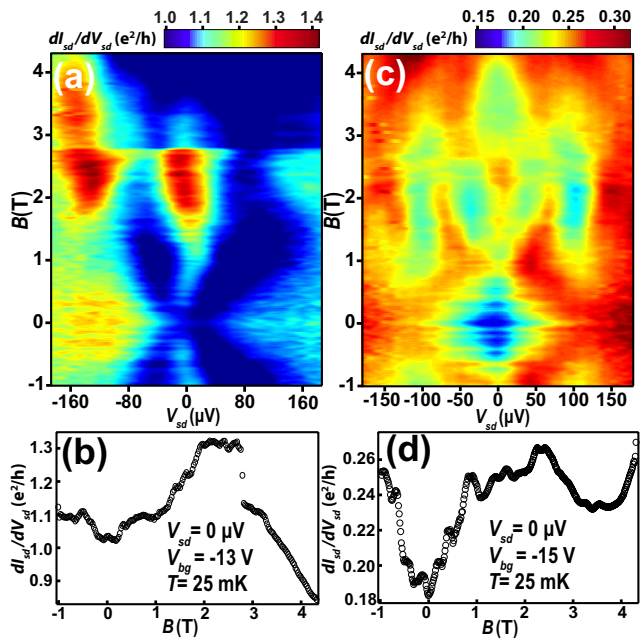}
\caption{\label{fig:4} \textbf{Zero-bias conductance plateau structures due to the transport through Majorana fermion states.} (a) Differential conductance as a function of source-drain bias voltage $V_{sd}$ and magnetic field $B$ applied perpendicularly to the substrate and thus to the InSb nanowire measured for the device at $V_{bg}=-13$ V and $T=25$ mK. (b) The same as in (a) but the differential conductance measured as a function of $B$ at $V_{sd}=0$ $\mu$V is plotted.  (c) The same as in (a)  but for $V_{bg}=-15$ V. (d) The same as in (b) but also for $V_{bg}=-15$ V. Here, enhanced zero-bias conductance plateau structures due to the transport through the Majorana fermion states in the InSb nanowire are again observed.}
\end{figure}

Zero-bias conductance plateau like structures have also been observed in other gate voltage regions. Figure~\ref{fig:4}a shows measurements of the differential conductance as a function of source-drain bias voltage $V_{sd}$ and applied magnetic field $B$ at a fixed back gate voltage of $V_{bg}=-13$ V, while Fig.~\ref{fig:4}c shows such measurements at $V_{bg}=-15$ V. Figures~\ref{fig:4}b and \ref{fig:4}d show the corresponding traces of the measurements at the zero source-drain bias voltage. Figures~\ref{fig:4}a and \ref{fig:4}b show that the zero-bias conductance has a value of $\sim 1.0  e^2/h$ at $B=0$ T and stay approximately at a constant value as $B$ increases. The zero-bias conductance starts to  increase quickly from $B\sim 1.3$ T and reaches a plateau value of $\sim 1.3  e^2/h$ at $B\sim 2$ T. The zero-bias conductance stays at the constant value until $B=2.8$ T and then jumps to a lower value as a results of a charge rearrangement (see Fig.~\ref{fig:4}a for the clear signature of the charge rearrangement). In fact, a plateau like zero-bias conductance structure can still be identified after the charge rearrangement. Figures~\ref{fig:4}c and \ref{fig:4}d show similar characteristics of the differential conductance. Here, the zero-bias conductance plateau can be identified to appear in a magnetic field range of $0.9-2.6$ T, although it is superimposed with fluctuations and a visible slope. In addition, the differential conductance measurements in the small $V_{sd}$ region are characterized by a gap at low magnetic fields, that nearly vanishes at the edge of the zero-bias conductance plateau and reopens as the magnetic field is continuously increased. This gap evolution behavior is similar to the one predicted in Ref.~\cite{SDS-PRB}. Here we should note that such a gap evolution structure is also visible in Figs.~\ref{fig:3}c and \ref{fig:4}a. However, its physical origin is not completely clear to us.  In Fig.~\ref{fig:4}c, we clearly see that the gap nearly vanishes again at the upper edge of the zero-bias conductance plateau and then reopens as the magnetic field is further increased and the system turns to be in a trivial topological superconductor phase.

In summary, we have studied a Nb-InSb nanowire quantum dot-Nb device by transport measurements and observed a zero-bias conductance plateau structure at finite perpendicularly applied magnetic fields. The height of the plateau has been found to depend on the back gate voltage and is most likely related to the phase difference between the two superconductor Nb contacts. At zero magnetic field, the Nb wrapped InSb nanowire segments is in a normal superconductor phase and our measurements show a tunable Josephson supercurrent in the current biased configuration and multiple Andreev reflection characteristics in the voltage biased configuration. Both the superconductor energy gap $\Delta_{Nb}$ of the Nb contact and the superconductor energy gap $\Delta_{InSb}$ of the Nb wrapped InSb nanowire segments have been identified from the measured multiple Andreev reflection characteristics. As the applied magnetic field increase, the entire InSb nanowire, including the two Nb wrapped segments and the middle quantum dot junction segment, turns to be in a nontrivial topological superconductor phase, supporting a pair of Majorana fermion states located at the two ends of the nanowire. Cooper pair transport through the InSb nanowire via Majorana fermion states leads to the observation of an enhanced zero-bias conductance plateau structure. As the magnetic field is further increased, the zero-bias conductance is found to drop from the plateau value to lower values. This could be interpreted as that the system has now been transformed to a trivial topological superconductor phase and thus shows the conventional Josephson junction transport characteristics.

\vspace{12pt}

\textbf{METHODS}

Our superconductor-semiconductor nanowire hybrid device is fabricated from the zincblende InSb segment of an InAs/InSb heterostructure nanowire. The heterostructure nanowires are grown on InAs(111)B substrates at 450 $^{\rm{o}}$C by metal-organic vapor-phase epitaxy in a two-stages process using aerosol Au particles as initial seeds. Growth of an InAs stem first replaces direct nucleation by a wire-on-wire growth process, effectively favoring a high yield of epitaxial top InSb nanowire segments. Contrary to most other III-V nanowires, our InSb segments are free of any extended structural defects and do not show tapering. For further details about the growth, structural, and basic field effect transistor properties of the InAs/InSb heterostructure nanowires, see Refs.~\cite{InSb-small} and \cite{Plissard12} and the references therein.

The grown InAs/InSb heterostructure nanowire is transferred to a 100 nm thick SiO$_2$ layer capped, degenerately doped, n-type Si substrate with predefined Ti/Au bonding pads and metal markers. Using an optical microscope, the wire position relative to the metal markers is recorded. The substrate with the heterostructure nanowire deposited is spin-coated with polymethylmethacrylate (PMMA) resist. Then, two 470 nm wide Ti/Nb/Ti (3 nm/80 nm/5 nm) superconductor contacts with a separation of 110 nm are defined on the InSb segment of the InAs/InSb heterostructure nanowire using electron beam lithography, sputtering and lift-off process. Here, we note that to remove the outside oxide layer of the InSb nanowire, a surface chemical retreatment in a (NH$_{4}$)$_{2}$S$_{x}$ solution is performed before metal deposition by sputtering. In addition to the two superconductor contacts, there is an extra Ti/Au metal layer on the back side of the chip which can serve as a global back gate. Figure~\ref{fig:1}a shows an SEM image of the device measured in this work. All electrical measurements reported in this work are performed in a $^3$He/$^4$He dilution refrigerator.


\textbf{Supplementary information} is available online at www.nature.com/nature.

\textbf{Acknowledgements}
This work was supported by the Swedish Research Council (VR), the Swedish Foundation for Strategic Research (SSF), the Chinese Scholarship Council (CSC), and the National Basic Research Program of the Ministry of Science and Technology of China (Nos.
2012CB932703 and 2012CB932700).

\textbf{Author Contributions}
M. T. Deng participated in the development of the idea for the experiment, led and participated in the device fabrication, performed the measurements and the data analysis, and participated in the developments of the physical concepts for the observation and in the writing of the paper. C. L. Yu fabricated the device and participated in the measurements.  G. Y. Huang performed numerical calculations. M. Larsson participated in the measurements. P. Caroff growed the InSb nanowires used for this work. H. Q. Xu had the idea for and initiated the experiment, participated in the data analysis, developed the physical concepts for the observation, wrote the manuscript, and supervised the whole project.

\textbf{Author Information} Reprints and permissions information is available at www.nature.com/reprints. The authors declare no competing
financial interests. Correspondence and requests for materials should be addressed to H. Q. Xu (hongqi.xu@ftf.lth.se).


\begin{thebibliography}{99}
%===============================================================================
\bibitem{MFreturns}
Wilczek, F. Majorana returns. \emph{Nature Physics} \textbf{5}, 614-618 (2009).

\bibitem{Franz10} Franz, M. Race for Majorana fermions. \emph{Physics} \textbf{3}, 24 (2010).

\bibitem{MFnearSuccess} Service, R.F. Search for Majorana fermions nearing success at last? \emph{Science} \textbf{332}, 193-195 (2011).


\bibitem{Beenakker11} Beenakker, C. W. J. Search for Majorana fermions in superconductors. \emph{arXiv:1112.1950}, 1-13 (2011).

\bibitem{MFinSolid}
Alicea, J. New directions in the pursuit of Majorana fermions in solid state systems. \emph{arXiv:1202.1293}, 1-36 (2012).

%------------------------------
%Majorana's original paper:
\bibitem{Majorana}
Majorana, E. Teoria simmetrica dell'elettronee del positrone. \emph{NUOVO CIMENTO} \textbf{14}, 171-184 (1937).
%------------------------------

%------------------------------
\bibitem{KitaevHybridized}
Kitaev, A.Y. Unpaired Majorana fermions in quantum wires. \emph{Physics-Uspekhi} \textbf{131}, 130-136 (2001).

%FQHE:

\bibitem{5-2FQHEasMF}
Nayak, C., Simon, S.H., Stern, A., Freedman, M. \& Das Sarma, S. Non-Abelian anyons and topological quantum computation. \emph{Reviews of Modern Physics} \textbf{80}, 1083-1159 (2008).

%Review article about Majorana:
\bibitem{NonAbelian}
Stern, A. Non-Abelian states of matter. \emph{Nature} \textbf{464}, 187-193 (2010).

%------------------------------

%------------------------------
%P-wave topological superconductor

\bibitem{ReadGreenPwave}
Read, N. \& Green, D., Paired states of fermions in two dimensions with breaking of parity and time-reversal symmetries and the fractional quantum Hall effect. \emph{Physical Review B} \textbf{61}, 10267-10297 (2000).
%------------------------------


%------------------------------
%S-wave superconductor coupled topological insulator.

\bibitem{FuKaneSwaveTi}
Fu, L. \& Kane, C. Superconducting Proximity Effect and Majorana Fermions at the Surface of a Topological Insulator. \emph{Physical Review Letters} \textbf{100}, 096407 (2008).
%------------------------------

%------------------------------
%S-wave S coupled 2D semiconductor.

\bibitem{JaySauMFinSemi}
Sau, J.D., Lutchyn, R.M., Tewari, S. \& Das Sarma, S. Generic New Platform for Topological Quantum Computation Using Semiconductor Heterostructures. \emph{Physical Review Letters} \textbf{104}, 040502 (2010).

\bibitem{AliceaMFin2DSemi}
Alicea, J. Majorana fermions in a tunable semiconductor device. \emph{Physical Review B} \textbf{81}, 125318 (2010).

\bibitem{SauPRB10}
Sau, J.D., Tewaru, S., Lutchyn, R.M., Stanescu, T.D. \& Das Sarma, S. Non-Abelian quantum order in spin-orbit-coupled semiconductor: Search for topological Majorana particles in solid-state systems. \emph{Physical Review B} \textbf{82}, 214509 (2010).
%------------------------------

%------------------------------
%S-wave S coupled 1D semiconductor.
\bibitem{DSarmaMFin1DSemi}
Lutchyn, R., Sau, J. \& Das Sarma, S. Majorana Fermions and a Topological Phase Transition in Semiconductor-Superconductor Heterostructures. \emph{Physical Review Letters} \textbf{105}, 077001 (2010).

\bibitem{OregHelicalMF}
Oreg, Y., Refael, G. \& von Oppen, F. Helical Liquids and Majorana Bound States in Quantum Wires. \emph{Physical Review Letters} \textbf{105}, 177002 (2010).

\bibitem{SDS-PRB}
Stanescu, T., Lutchyn, R. \& Das Sarma, S. Majorana fermions in semiconductor nanowires. \emph{Physical Review B} \textbf{84}, 144522 (2011).

%------------------------------
%Majorana to quantum computing:

\bibitem{MF1DNetwork}
Alicea, J., Oreg, Y., Refael, G., von Oppen, F. \& Fisher, M. P. a. Non-Abelian statistics and topological quantum information processing in 1D wire networks. \emph{Nature Physics} \textbf{7}, 412-417 (2011).


\bibitem{FlensbergMFQD2}
Leijnse, M. \& Flensberg, K. Quantum Information Transfer between Topological and Spin Qubit Systems. \emph{Physical Review Letters} \textbf{107}, 210502 (2011).


%------------------------------

\bibitem{Giant-gfactor}
Nilsson, H. A., Caroff, P., Thelander, C., Larsson, M., Wagner, J. B., Wernersson, L.-E., Samuelson, L. \& Xu, H. Q. Giant, level-dependent g factors in InSb nanowire quantum dots. \emph{Nano Letters} \textbf{9}, 3151-3156 (2009).

\bibitem{Henrik-PRL}
Nilsson, H. A., Karlstr\"{o}m, O., Larsson, M., Caroff, P., Pedersen, J. N.,  Samuelson, L., Wacker, A., Wernersson, L.-E. \& Xu, H. Q. Correlation-Induced Conductance Suppression at Level Degeneracy in a Quantum Dot. \emph{Physical Review Letters} \textbf{104}, 186804(2010).

\bibitem{Henrik-Super}
Nilsson, H. A., Samuelsson, P., Caroff, P. \& Xu, H. Q. Supercurrent and multiple Andreev reflections in an InSb nanowire Josephson junction. \emph{Nano Letters} \textbf{12}, 228-233 (2012).

\bibitem{Hc2OfNb}
Bose, S., Raychaudhuri, P., Banerjee,R., \& Ayyub, Pushan., Upper critical field in nanostructured Nb: Competing effects of the reduction in density of states and the mean free path. \emph{Physical Review B} \emph{74}, 224502 (2006).

\bibitem{Doh05} Doh, Y.-J., van Dam, J. A., Roest, A. L., Bakkers, E. P. A. M., Kouwenhoven, L. P. \& De Franceschi, S. \textit{Science} \textbf{309}, 272-275 (2005).


\bibitem{Xiang06} Xiang, J., Vidan, A., Tinkham, M., Westervelt, R. M. \&  Lieber, C. M. Ge/Si nanowire mesoscopic Josephson junctions \textit{Nature Nanotech.} \textbf{1}, 208-213 (2006).

\bibitem{Dam06} van Dam, J. A., Nazarov, Y. V., Bakkers, E. P. A. M., De Franceschi, S. \& Kouwenhoven, L. P. Supercurrent reversal in quantum dots. \emph{Nature} \textbf{442}, 667-670 (2006).

\bibitem{Nb-InGaAs-PRB}
Deon, F. et al. Proximity effect in a two-dimensional electron gas probed with a lateral quantum dot. \emph{Physical Review B} \textbf{84}, 100506 (2011).


\bibitem{Nb-Detpth}
Maxfield, B. \& McLean, W.L. Superconducting penetration depth of niobium. \emph{Phys Rev} \textbf{139}, 5A(1965).

%------------------------------
%Hybridized Majorana:


%===============================================================================

\bibitem{InSb-small}
Caroff, P. et al. High-quality InAs/InSb nanowire heterostructures grown by metal-organic vapor-phase epitaxy. \emph{Small} \textbf{4}, 878-882 (2008).

\bibitem{Plissard12} Plissard, S. R.,Slapak, D. R., Verheijen, M. A., Hocevar, M.; Immink, G. W. G., van Weperen, I., Nadj-Perge, S., Frolov, S. M., Kouwenhoven, L. P. \& Bakkers, E. P. A. M. From InSb Nanowires to Nanocubes: Looking for the Sweet Spot. \emph{Nano Letters} in press (2012).


\end{thebibliography}
\end{document}

\title{Supplementary Information\\
Observation of Majorana Fermions in a Nb-InSb Nanowire-Nb Hybrid Quantum Device\\
Theoretical Model and Simulation Results}

\author{M. T. Deng}
\affiliation{Division of Solid State Physics, Lund University, Box 118, S-221 00 Lund, Sweden}
\author{C. L. Yu}
\affiliation{Division of Solid State Physics, Lund University, Box 118, S-221 00 Lund, Sweden}
\author{G. Y. Huang}
\affiliation{Division of Solid State Physics, Lund University, Box 118, S-221 00 Lund, Sweden}
\author{M. Larsson}
\affiliation{Division of Solid State Physics, Lund University, Box 118, S-221 00 Lund, Sweden}
\author{P. Caroff}
\affiliation{I.E.M.N., UMR CNRS 8520, Avenue Poincar\'{e},  BP 60069, F-59652 Villeneuve d'Ascq, France}
\author{H. Q. Xu}
\email[Corresponding author: ]{hongqi.xu@ftf.lth.se}
\affiliation{Division of Solid State Physics, Lund University, Box 118, S-221 00 Lund, Sweden}
\affiliation{Department of Electronics and Key Laboratory for the Physics and Chemistry of Nanodevices, Peking University, Beijing 100871, China}

\date{March 27, 2012}

\maketitle

%\section{Theoretical Model and simulation results}

\begin{figure}[hb]
\includegraphics[width=8.5cm]{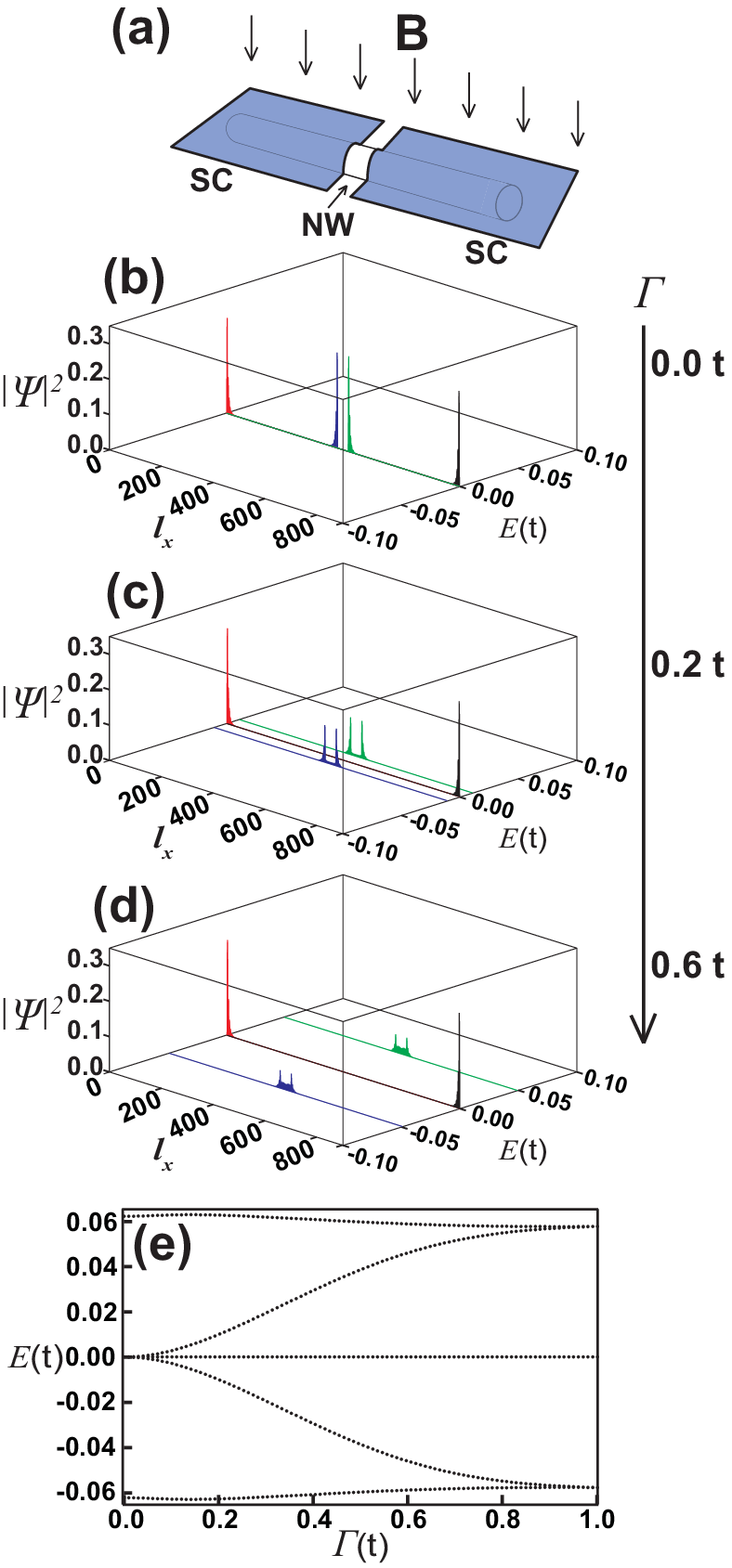}
\caption{\label{fig:S1} (a) Schematic of the model system consisting of a semiconductor nanowire (NW), with a strong Rashba SOI,  and two superconductor (SC) contacts (grey regions). The nanowire is placed along the $x$ axis. Due to the proximity effect, the nanowire segments wrapped by the superconductor contacts are in a superconducting state with a superconductor energy gap $\Delta^*$. These two superconducting segments are connected to the normal semiconductor segment via tunnel couplings $\Gamma$. The magnetic field $B$ is applied perpendicularly to the nanowire along the $z$ axis. (b) to (d) Wave function probability distributions along the nanowire for the first few low energy states calculated for $\Gamma=0$, $0.2t$, and $0.6t$. The nanowire is modeled by 901 lattice sites with the normal semiconductor segment represented by 42 sites in the middle section. The other parameters used in the calculations are $\Delta^*=0.2 t$,  $V_z=0.5 t$, and $\alpha_0=1$. (e) Energies of the low energy states calculated for the same system as in (b) to (d) as a function of $\Gamma $. The two quasi-particle states emerge from the hybridization of the two interacted Majorana fermion states at a finite $\Gamma$ and move apart in energy with increasing $\Gamma$.}
\end{figure}

According to the experimental setup as sketched in Fig.~\ref{fig:S1}, we establish a semiconductor nanowire model which consists of three segments: two proximity effect induced superconducting segments located under two superconductor contacts and a middle normal semiconductor junction segment. The entire nanowire possesses a strong spin-orbit interaction (SOI) and the normal semiconductor segment is connected to the two superconducting segments via tunnel couplings in order to take into account the weak Schottky effect induced by the two superconductor contacts to the nanowire. Furthermore, we assume a magnetic field is applied perpendicularly to the nanowire as indicated in Fig.~\ref{fig:S1}. Our semiconductor nanowire model resembles the superconductor-semiconductor-superconductor heterostructure system of Lutchyn \emph{et al.} \cite{SDS-PRL1}, but with two major differences. (i) The phase difference of superconducting order parameter of the left and right segment is unknown and is taken to be a constant in the calculations presented below. Here, we concentrate on the energy spectrum rather than Josephson current. (ii) The couplings between the normal semiconductor segment and the two superconducting segments are equal but tunable, i.e., we consider a simplest system of a symmetric double barrier type.

The Hamiltonian of our model system is constructed by assuming that the nanowire is oriented along the $x$ axis, the SOI is of the Rashba type, and the electric and magnetic field are applied along the $z$ direction. Under the Landau gauge, the Hamiltonian can be written in an one-dimensional  tight-binding form as
\begin{eqnarray}
(\mathcal{H}_{BdG})_{n,m}&=&[(2t-\mu)\tau_{z}+V_{z}\sigma_{z}+\Delta_{n}\tau_{x}]\delta_{n,m}\nonumber\\
&+&(-t_{n}+i\alpha_{n}\sigma_{y})\tau_{z}\delta_{n,m-1}+(-t_{n}-i\alpha_{n}\sigma_{y})\tau_{z}\delta_{n,m+1}.\nonumber
 \end{eqnarray}
In the equation above, indices $m$ and $n$ run through the lattice sites in the model. Site dependent parameters $t_{n}$ are the hopping integrals with $t_{n}=\Gamma$ at the two interfaces between the semiconductor and two superconducting segments and $t_{n}=t$ otherwise. $\Delta_{n}=\Delta^*$
in the left superconducting segment, $\Delta^* e^{i\phi}$ in the right superconducting segment, and 0 in the middle normal semiconductor segment, where $\Delta^*$ is the proximity effect induced superconductor energy gap in the two superconductor contacted nanowire segments, and $\phi$ is the phase difference of the
two superconducting nanowire segments and is set to zero in the calculations presented here. $\alpha_{n}=t_n \alpha_0$ is the Rashba SOI energy with $\alpha_0$ being the dimensionless parameter describing the Rashba SOI strength. $V_{z}=\frac{1}{2}g\mu_{B}B$ is the Zeeman energy, where $g$ the effective $g$ factor, $\mu_{B}$ the Bohr magneton, and $B$ the magnetic field. Finally, $\sigma_{i}$ and $\tau_{i}$ are Pauli matrices which operate on spin space and particle-hole space, respectively.

The above model has been employed to explore the energy spectrum evolution with increasing coupling
$\Gamma$. A special emphasis is placed on the properties of the zero energy Majorana fermion states for the $\mu=0$ case. Figures~\ref{fig:S1}(b)-\ref{fig:S1}(d) show the results of the calculations for the system, modeled by 901 lattice sites with the middle normal semiconductor segment represented by 42 sites, at $\Delta^*=0.2 t$, $V_z=0.5 t$, and $\alpha_0=1$, and at three different coupings $\Gamma$. In these figures, the wave function probability distributions are plotted for a few low energy states. At $\Gamma=0$, we find two pairs of zero energy Majorana fermion states with each pair being localized at the two ends of a proximity effect induced superconducting nanowire segment. As $\Gamma$ becomes finite, the two Majorana fermion states near the middle normal semiconductor segment interact, leading to the creation of a pair of normal fermion states--a quasi-particle state and a quasi-hole state--with energies located symmetrically around the zero~\cite{Helical-in-1D, pu44.131, prl107.236401,np7.412,arxiv1202.1293,prb84.094505}. However, the other two Majorana fermion states seen at $\Gamma=0$ remain intact and the entire nanowire behaves as a nontrivial topological superconductor system. At $\Gamma=0.6t$, the two normal fermion states are seen to move further apart in energy, while the two Majorana fermion states still remain unchanged. A summary of the results of the calculation is shown in Figure~\ref{fig:S1}(e) where the continuous energy evolutions of a few low energy states with increasing $\Gamma$ are presented.

In conclusion, when two proximity effect induced superconducting semiconductor nanowires in a nontrivial topological superconductor phase are brought to be coherently coupled via a normal semiconductor nanowire segment, the two Majorana fermion states localized near the normal semiconductor nanowire segment will be annihilated to create a pair of normal fermion states--a quasi-particle and a quasi-hole fermion state--and the entire nanowire will turn to be a nontrivial topological superconductor system with a pair of Majorana fermions localized at the two ends of the entire nanowire. Cooper pairs can transport through the semiconductor nanowire via excitation of the pair of Majorana fermions \cite{prb85.085415}, leading to an enhancement of the zero-bias conductance measured in the experiment reported in the main article.

\textbf{Acknowledgements}
This work was supported by the Swedish Research Council (VR), the Swedish Foundation for Strategic Research (SSF), the Chinese Scholarship Council (CSC), and the National Basic Research Program of the Ministry of Science and Technology of China (Nos.
2012CB932703 and 2012CB932700).